# Magnetic Manipulation of Spatially Confined Multiferroic Heuslers by Martensitic Microstructure Engineering

Milad Takhsha[1*], Vipin Kumar Singh[2], Julian Ledieu[2], Simone Fabbrici[1*], Francesca Casoli[1*], Francesco Mezzadri[3], Michal Horký[4], Vincent Fournée[2], Vojtěch Uhlíř[4,5], Franca Albertini[1]

[1]IMEM-CNR, Parco Area delle Scienze 37/A, 43124 Parma, Italy

[2]Institut Jean Lamour, UMR7198 CNRS-Nancy-Université de Lorraine, Campus ARTEM –

2 allée André Guinier, BP 50840, 54011 Nancy, France

[3]Dipartimento di Scienze Chimiche, della Vita e della Sostenibilità Ambientale, Parco Area

delle Scienze 17/A, 43124 Parma, Italy

[4] CEITEC BUT, Brno University of Technology, Purkyňova 123, 61200 Brno, Czech Republic

[5]Institute of Physical Engineering, Brno University of Technology, Technická 2, 61669 Brno,

Czech Republic

*Corresponding authors:

M.T.: milad.takhsha@imem.cnr.it,

S.F.: simone.fabbrici@imem.cnr.it

F.C.: francesca.casoli@imem.cnr.it

**Abstract**

Magnetic shape-memory (MSM) Heuslers show a strong coupling between magnetic and structural characteristics, evidencing a correlation between magnetic, thermal, and mechanical properties through a magnetostructural martensitic transformation. This functional aspect makes MSM Heuslers promising for integration into smart micro/nanodevices, including sensors, energy harvesters, and actuators. Controlling the martensitic microstructure, which determines the magnetic characteristics, is among the key points for optimization of the magnetic functional properties of these materials at different length scales. Here, we report a strategy for manipulating the magnetic properties of spatially confined epitaxial Ni-Mn-Ga films grown on Cr(001)//MgO(001) by twinning configuration engineering in the low-temperature ferromagnetic (martensitic) phase. We show how the twinning configurations in the continuous films and the micropatterned structures can be switched from Y-type (showing negligible magnetic stray field) into X-type (presenting significant magnetic stray field) by a post-annealing process. Advanced characterization techniques enable us to analyze the atomic structure and the surface quality of the annealed samples and to disentangle the "twin-switching" phenomenon. The martensitic microstructure engineering reported in this study introduces a simple method for promoting the magnetic stray-field contribution at the surface of Ni-Mn-Ga epitaxial thin films and micropatterns initially showing a negligible magnetic stray field.

**Keywords:** Magnetic shape-memory alloys; Multiferroics; Heusler alloys; Martensitic phase transformation; Twin boundaries; Lithography patterning; Surface analysis.

# 1. Introduction

Magnetic shape-memory (MSM) Heuslers are multiferroic materials showing ferroelastic and ferromagnetic properties [1]. They exhibit strong coupling between magnetic and structural characteristics, evidencing a correlation between magnetic, thermal, and mechanical properties through a magnetostructural phase transformation [2-4]. In particular, MSM Heusler thin films are of special interest for the integration into smart micro/nanodevices such as sensors, energy harvesters, and actuators with various promising applications [5-8]. Recently, the successful epitaxial growth of MSM Heusler thin films on silicon substrates using $SrTiO_3$ buffer layers has facilitated the integration of these materials into micro/nanoelectronics and micro/nanomachining technologies based on silicon [9]. Martensitic microstructure dictates the magnetic characteristics of these films; therefore, controlling the twinning configurations at different length scales is the key to the optimization of the material functional magnetic properties.

Upon temperature variation, MSM Heuslers undergo a phase transformation between a high-temperature high-symmetry phase (austenitic) and a low-temperature low-symmetry phase (martensitic). The martensitic phase consists of self-assembled structural elastic domains (twin variants), which accommodate differently oriented cells in complex hierarchical patterns [10-12]. Epitaxial thin films of MSM Heuslers such as Ni-Mn-X (Ga, In, Sn, Al) and their Co-doped compounds [13-16], having film(001)//underlayer(001)//substrate(001) crystallographic relationship, consist of either both or one of the differently oriented hierarchical twinning configurations in the martensitic phase called X-type and Y-type [14,17-23]. The mechanism leading to the selection of each of the two configurations is not yet well understood. However, it is assumed to be based on stress conditions [24-28] across the phase transformation where *b*-axis of the martensitic cells (in the austenitic setting) lies in the plane of the film to construct the X-type or out of the plane of the film to construct the Y-type (**Figure 1**). If the film is ferromagnetic in the martensitic phase, the coupling between magnetic and structural degrees of freedom gives rise to two different magnetization patterns depending on the twinning configuration. For X-type twinning configurations, the easy-magnetization axis (*c*-axis) of the cells alternates between out-of-plane and in-plane, while for Y-type twinning configurations, the easy-magnetization axis of the cells alternates in the plane of the film along the MgO diagonal [14,17]. Therefore, different arrangements of twinning configurations give rise to thin films with various magnetic stray-field contributions at the surface (Figure 1). In previous works, we have reported a few strategies for engineering X-type and Y-type in epitaxial Ni-Mn-Ga films [17,18,27,29]. Controlling these twinning configurations in a film is a subtle process depending on many parameters such as composition, thickness, underlayer/substrate, growth conditions, and post-growth treatments.

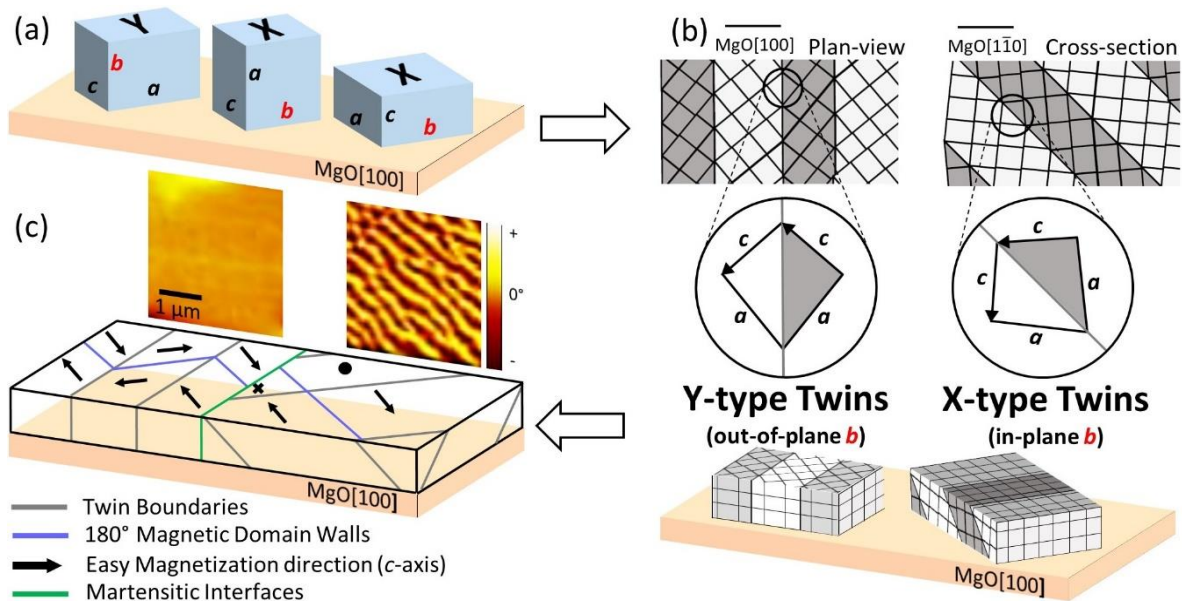

*Figure 1: (a) 3D scheme of the three principal martensitic cell orientations with respect to the substrate plane. (b) Simplified representation of the martensitic cell arrangement in Y-type and X-type twins; (c) 3D slice of the film showing simplified scheme of Y-type and X-type twins (grey lines), martensitic interfaces (green lines), 180° magnetic domain walls (purple lines) and the easy magnetization direction in the martensitic cells (black arrows) across the twin boundaries. The resultant magnetic contribution from X-type and Y-type on the surface is shown by the MFM micrographs.*

In this study, we report a simple strategy for switching the entire martensitic configuration of Ni-Mn-Ga films, epitaxially grown on Cr(001)//MgO(001) with dominant Y-type twins showing negligible magnetic stray field contribution at the surface. We show that the magnetic stray field contribution at the surface can be significantly promoted by switching the twinning configurations of the pristine films into X-type as a result of a post-annealing treatment while maintaining the high quality of the surface. With the help of advanced characterization techniques, we evaluate the evolution of the microstructure of the sample before and after the annealing process, and we assess the surface characteristics down to the atomic scale. We provide evidence showing that this simple twinning configuration engineering approach works perfectly both for continuous films and spatially confined micropatterns. Therefore, without the need for complicated effort, the twinning configurations can be switched into X-type while maintaining the high surface quality of the samples giving rise to considerable magnetic stray field contribution. Finally, we discuss the observed "switching of twins" by introducing a phenomenological hypothesis, based on which, the directions of internal stress in the epitaxial films contribute to the selection of the twinning configurations.

**2. Experimental**

2.1. Growth of Thin Films and Annealing

Epitaxial Ni-Mn-Ga films (200-nm thick) were grown by radio-frequency sputtering technique on Cr (14-50 nm thick) underlayer//MgO(001) substrate at elevated temperature ($T_{growth}$ = 623 K). The temperature was ramped up ($T_{ramp}$ ≈ 5 K/min) to 623 K and stabilized in high vacuum (P = $10^{-6}$ Pa); then the chamber pressure was set to 1.5 Pa Ar pressure. The deposition rates were set to 0.1 nm/sec and 0.02 nm/sec for Ni-Mn-Ga and Cr, respectively. After the deposition accomplishment, the films were left overnight in the growth chamber to gradually cool down to room temperature in a high vacuum (P = $10^{-6}$ Pa), giving rise to $Ni_{54.35}Mn_{20.52}Ga_{25.13}$ (uncertainty ≈ 1%) films having Ni-Mn-Ga(001)[100]//Cr(001)[100]//MgO(001)[110] crystallographic relationship.

The annealing process was performed in high vacuum (P = $10^{-6}$ Pa) and ex-situ by gradually ramping up the room-temperature pristine films to 623 K ($T_{ramp}$ ≈ 5 K/min), keeping the samples at that temperature for 1 hour and cooling the films back to room temperature, gradually. No considerable variation in composition was detected after the annealing process.

### 2.2. Structural, Microstructural, and Magnetic Characterizations

The structural characterization of the samples was performed using a Rigaku Smartlab XE diffractometer making use of monochromatic Cu-K$_{\alpha 1}$ radiation, obtained with a Ge (220) double bounce crystal. An HyPix detector was operated in a 1D mode to collect reciprocal space maps in *2θ*-step/*ω*-scan geometry.

The magnetic micrographs were measured at room temperature using a Dimension 3100 scanning probe microscope with a MESP-V2 magnetic tip in the interleave mode. Compositional measurements and scanning electron microscopy (SEM) micrographs were obtained using a FEI-VERIOS 460 L SEM equipped with an energy-dispersive spectroscopy analytical detector and low-voltage solid-state backscattered electron (BSE) detector. Polarized-light images were captured using a magneto-optical Kerr microscope (Evico Magnetics).

Magnetic characterization of the samples as a function of the magnetic field and the electrical resistance measurements over temperature (200–400 K ($dT/dt$ = 0.5 K.min$^{-1}$)) were performed using a Cryogen-free physical property measurement system – Quantum Design, VERSALAB. The magnetic field (up to $\mu_0 H$ = |1.5| T) was applied in the plane of the film along MgO[100], MgO[110], and MgO[010] at room temperature and the corresponding magnetic moment of the samples was collected using the vibrating sample magnetometer (VSM) mode of the instrument, whereas the electrical measurements were obtained using the electrical transport mode of the instrument.

### 2.3. Surface Study

The thin film was mounted on a tantalum plate and inserted in a multi-chamber ultra-high vacuum (UHV) system in which the base pressure is $10^{-8}$ Pa or lower. A clean surface could be prepared by repeated sputtering ($Ar^+$, 1.5 kV, 15 min) and annealing cycles (up to 673 K, 30 min). The temperature was measured using an optical pyrometer with an emissivity that was determined equal to 0.45 using a bichromatic pyrometer. The surface cleanliness was verified by X-ray photoemission spectroscopy (XPS) using a non-monochromatized Mg Kα source and the near-surface composition was deduced from the intensities of the Ga *2p*, Ni *2p,* and Mn *2p* core-levels (not shown). The surface crystallographic structure was monitored by low-energy electron diffraction (LEED), which gives information on the quality of the long-range surface order and the surface crystal symmetry. The morphology and the local atomic arrangement at the surface were investigated using variable temperature scanning tunneling microscopy (VT-STM). STM images were processed with the WSxM software [30].

### 2.4. Microfabrication

Microfabrication was performed using the UV lithography process. The details of the process have been published elsewhere [29]. The sample was coated by a negative photoresist on a SUSS WETBENCH LabSpin coater and exposed to UV light in a SUSS-MA8 mask aligner. Then, the sample was developed in a standard chemical solution and went through a dry-etch process by Ar ions in a PlasmaPro 80 Oxford Instrument. Finally, the residuals of the photoresist were washed off. The annealing process of the sample was performed at 623 K for 60 min in a UHV chamber (P = $10^{-6}$ Pa).

### 3. Results

### 3.1. Structural, Magnetic and Microstructural Characterizations

The martensitic transformation of the thin film sample before and after the annealing process was traced by electrical resistance measurement as a function of temperature [29] (**Figure 2**a). Different steps of the phase transformation of the sample are highlighted in the figure, where the cubic austenitic phase, low-symmetry martensitic phase, and the temperature region related to the phase coexistence are labeled. The sample shows a negligible variation in the phase transformation temperature after the annealing process. Through this temperature-dependent characterization, we have confirmed that the sample is in the martensitic phase at room temperature, so the subsequent measurements were performed at room temperature.

The BSE micrographs give insight into the large-scale distribution of the X-type and Y-type twinning configurations in the sample before and after the annealing process. The large-scale micrographs in Figure 2b show the typical channeling contrast for Y-type (bright) and X-type (dark) twins [27] suggesting the dominance of Y-type twins; in addition, the high-resolution micrograph in the panel verifies the presence of these twins characterized and labeled by their typical relative orientations [17]. Figure 2c shows the micrographs of the same sample after the annealing treatment; no bright contrast is detectable anymore, suggesting the absence of Y-type twins; instead, X-type twins cover the entire scanned area.

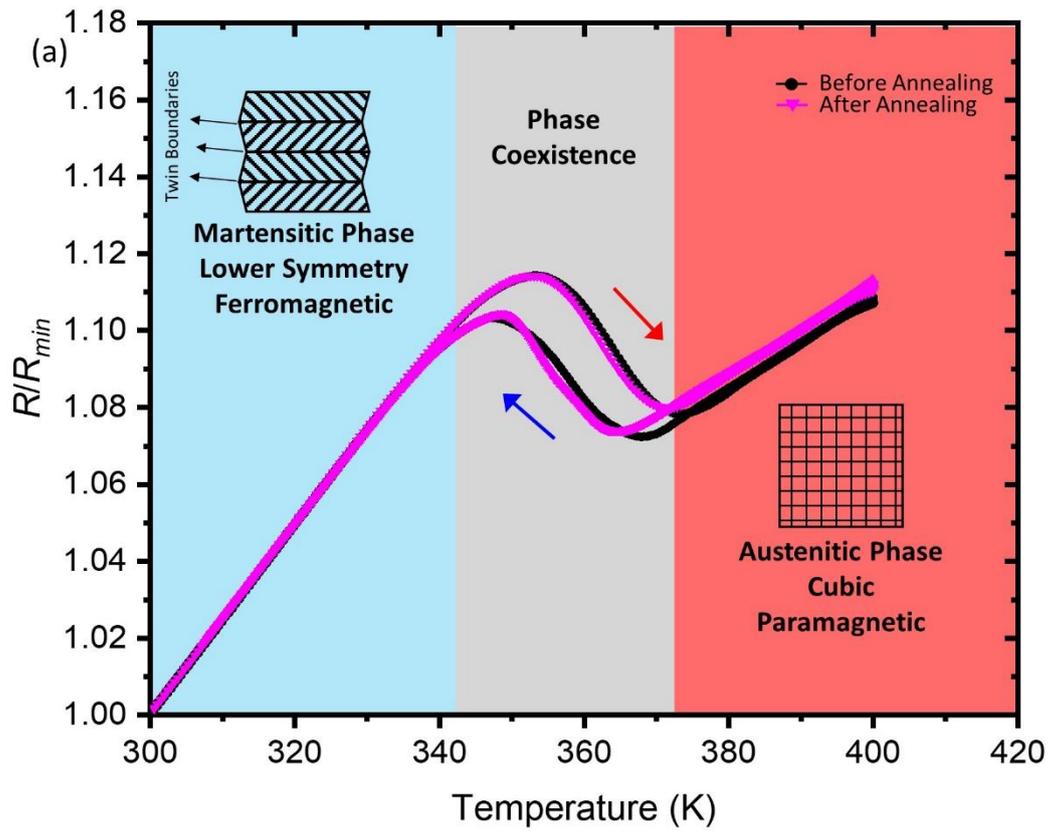

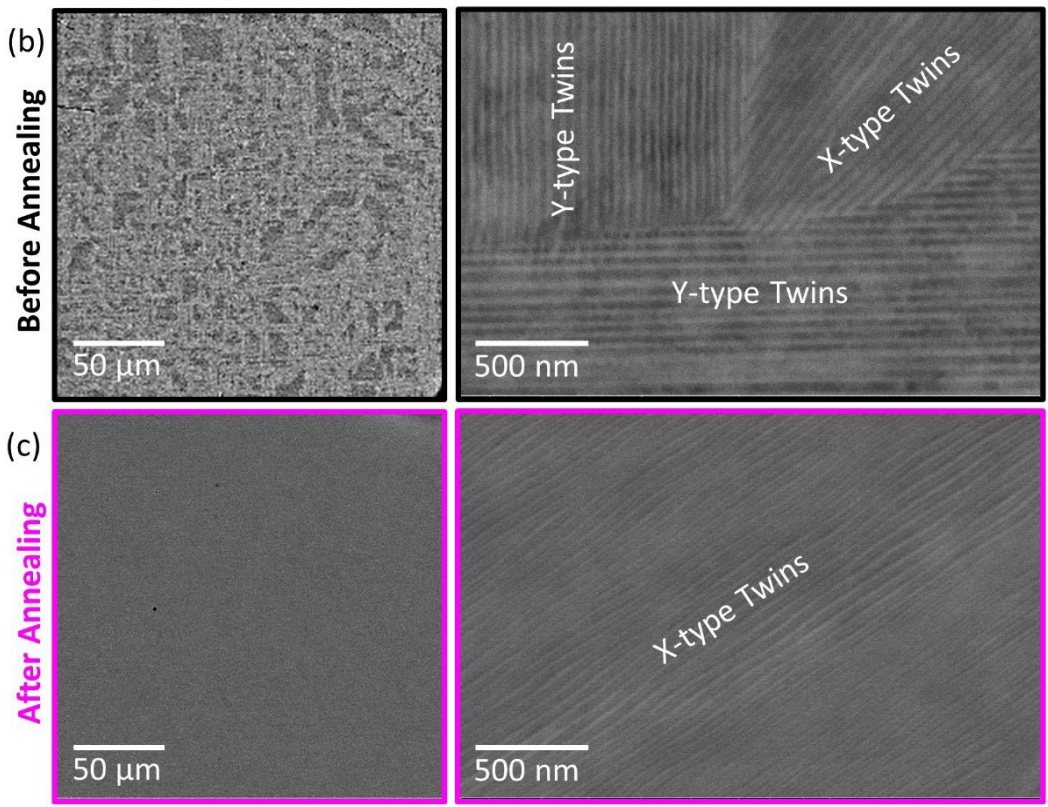

*Figure 2: Relative resistance curves as a function of temperature for the sample before (black circles) and after (pink triangles) the annealing process. BSE micrographs of the sample (b) before and (c) after the annealing process.*

The angular X-ray maps measured for the thin film before and after the annealing treatment are shown and compared in **Figure 3**. The 2θ/Δω maps, in this specific case, give us insight into the martensitic cells that are the building blocks for the martensitic twinning configurations (X-type and Y-type). As shown in Figure 1a, if we simplify the orientations of the martensitic monoclinic cells with respect to the substrate, considering them as pseudo-orthorhombic in the austenitic setting, we will have three principal orientations with *a*-axis, *b*-axis, or *c*-axis of the cells out of the plane of the film. To access these cell parameters, the 2θ/Δω angular dataset in the out-of-plane diffraction configuration was acquired to match the diffraction condition of those Ni-Mn-Ga crystallographic planes that are approximately parallel to the film plane. Therefore, the polar angle (χ) was set to zero degrees so that the diffraction from the Ni-Mn-Ga lattice planes was coherent with the substrate. The presence of different diffraction peaks with various intensities in the 2θ/Δω angular maps confirms the coexistence of different martensitic cells. For the sake of simplicity, we have highlighted the obtained peaks based on the aforementioned three principal martensitic cells having the *a*-axis, *b*-axis, or *c*-axis out of the plane of the film (labeled in Figure 3a, Figure 3b). In particular, the obtained out-of-plane maps reveal reflections at 2θ = 60.2°, 2θ = 64.0°, 2θ = 68.3°, which can be assigned to the (400), (040), and (004) reflections arising from *a*, *b*, and *c* lattice planes of the martensitic cells [29]. In addition, there is a high-intensity reflection at 2θ = 64.4°, which is assigned to the (002) reflection arising from the Cr underlayer lattice planes [17]. Comparing the relative intensity of the obtained reflection peaks for the sample before and after annealing in Figure 3a and Figure 3b gives us insight into the relative abundance of each of the three principal martensitic cells having *a*-axis, *b*-axis, or *c*-axis out of the plane of the film. The pristine sample in Figure 3a shows a relatively high-intensity reflection for the *b* lattice planes of the martensitic cells, suggesting the abundance of the building blocks for Y-type twins in this sample. The sample after the annealing process shows a completely different intensity pattern (Figure 3b): the reflection for the *b* lattice planes has disappeared, suggesting the absence of the building blocks for Y-type twins. Based on the obtained data, one can conclude that the annealing process has eliminated the population of the cells constituting Y-type twins in the sample.

Details on the annealing-induced magnetic evolution of the sample were obtained by magnetic force microscopy shown in Figure 3c and Figure 3d. The micrograph after the annealing process clearly shows the dominance of the out-of-plane magnetic domains in X-type twins arising from the perpendicular anisotropy contribution of the martensitic cells having the *c*-axis out of the plane of the film. The dominant zero-signal areas in the magnetic micrograph obtained for the sample before annealing can be ascribed to the Y-type twins with the *c*-axis of the martensitic cells in the plane of the film, therefore having negligible perpendicular anisotropy contribution [14,17,21,22,27].

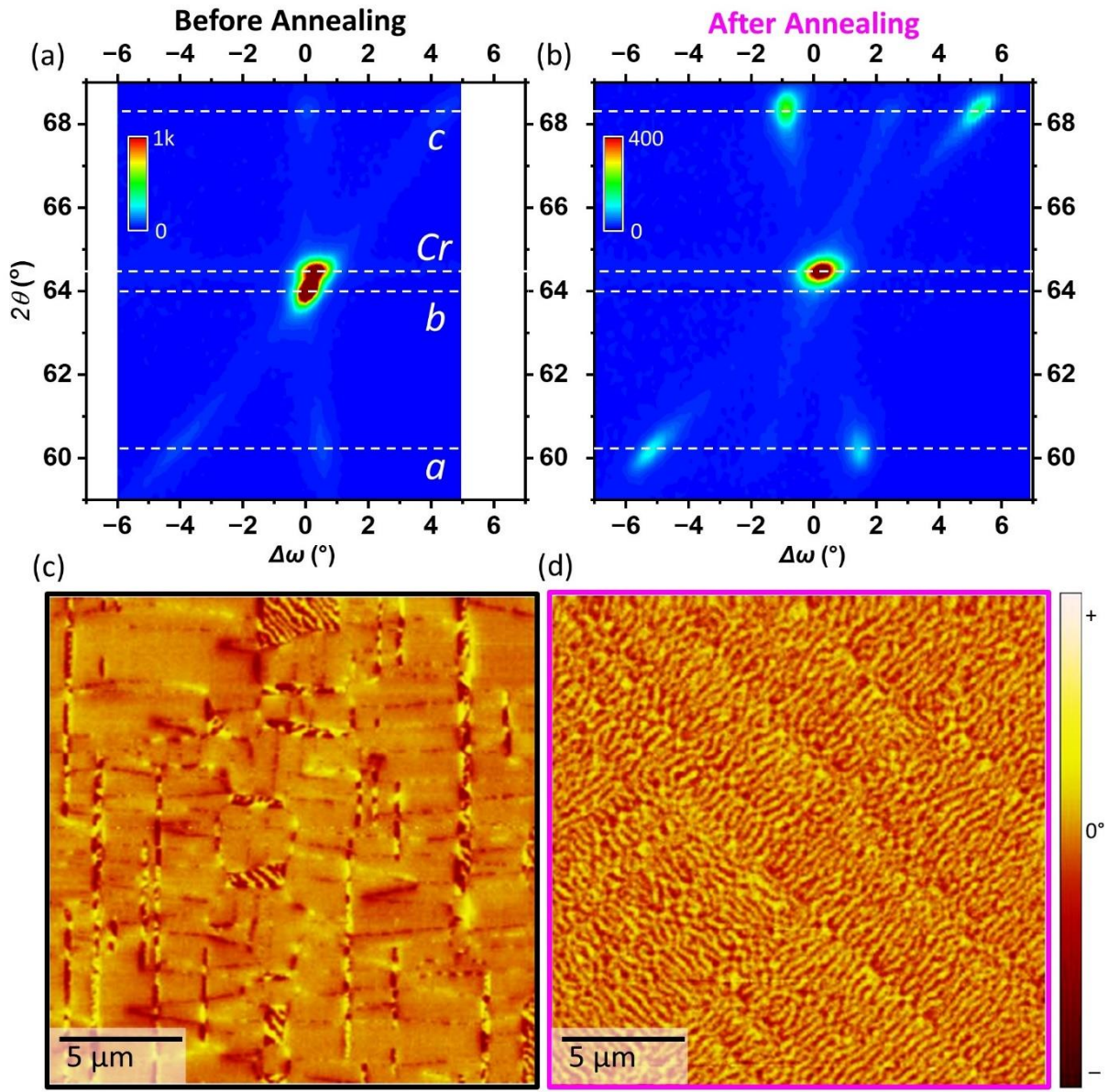

*Figure 3: 2θ/Δω angular maps of the sample (a) before and (b) after the annealing process in the martensitic phase at room temperature. MFM micrographs of the sample (c) before and (d) after the annealing process in the martensitic phase at room temperature.*

The unique arrangement of the martensitic cells before and after the annealing process gives rise to completely different magnetic characteristics of the sample. Figure 4 shows the magnetization loops of the sample before and after the annealing process as a function of an external magnetic field up to ±1.5 T that was applied along three in-plane directions of the sample at room temperature. Among these three applied magnetic field directions, only MgO[110] is parallel to one of the three easy-magnetization directions occurring in X-type twins and one of the two easy-magnetization directions occurring in Y-type twins (Figure 1). The magnetization loops in Figure 4a show magnetization jumps in both the first and the third quadrants at around 50 mT when we apply the magnetic field along MgO[100] and MgO[010], which are among the magnetically hard directions of the system (Figure 1). The loops measured for these two directions also show slightly lower remanence and coercivity compared to the measured loop when we applied the magnetic field along MgO[110]. The same magnetization measurements obtained for the sample after the annealing process (Figure 4b) show no trace of the magnetization jump and no considerable variation in the remanence and coercivity when we applied the magnetic field along MgO[100], MgO[010], or MgO[110]. The other noticeable point is the significant reduction of the remanence and coercivity of the loops (76% and 56%, respectively) obtained for the sample after the annealing process.

In one of our recent works [21], we have described the magnetization process of Ni-Mn-Ga epitaxial films in the martensitic phase where the spatial arrangement of magnetocrystalline anisotropy related to the martensitic twinning configurations plays an important role. The micromagnetic simulations also describe the magnetization jumps in the magnetization loops of the samples as having Y-type twinning configurations. These typical jumps occur when the magnetization reversal proceeds with the formation and expansion of magnetic domains, passing around the zero field in a closed-flux domain configuration [21].

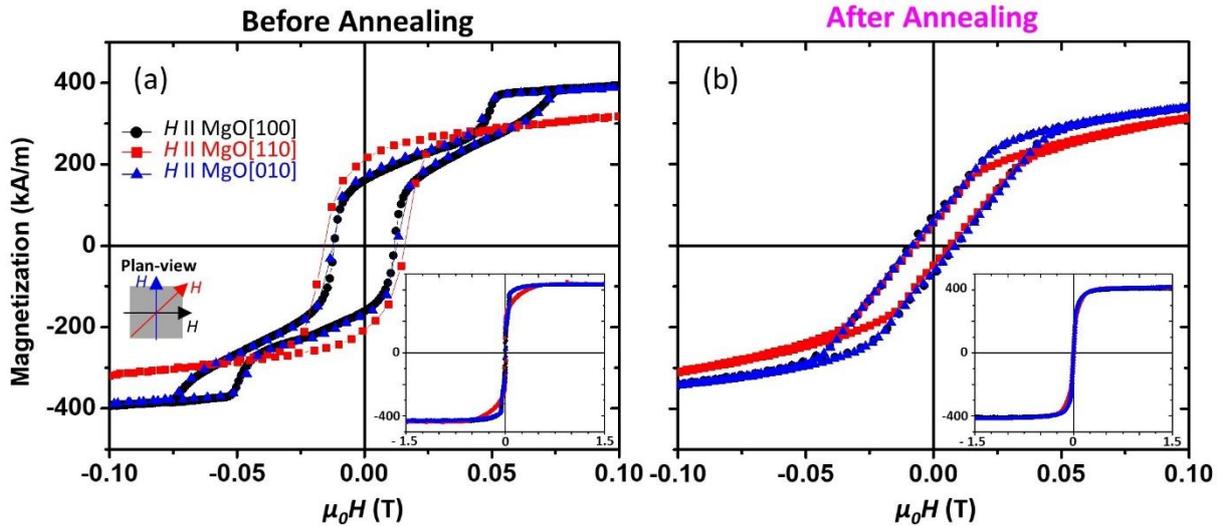

*Figure 4: Magnetization reversal curves of the sample at room temperature (a) before and (b) after the annealing process, applying the magnetic field along the different in-plane orientations of the sample shown by color code and symbols.*

3.2. Surface Study

To deeply analyze the atomic structure, superstructure, and surface characteristics of our films after the annealing process, a similarly grown sample was inserted into an ultra-high vacuum chamber equipped with advanced surface characterization techniques, including STM and temperature-dependent LEED.

The LEED pattern of the surface was monitored upon cooling after the annealing process. **Figure 5**a shows a typical LEED pattern of the sample at 50 eV primary electron beam energy at 348 K as measured by the thermocouple attached close to the sample holder. Sharp diffraction spots with a low background intensity indicate a clean surface with long-range crystallographic order. The pattern is quadratic with p(1x1) symmetry consistent with a cubic austenitic structure with (001) surface orientation. A scheme of the austenitic unit cell and its orientation with respect to the MgO substrate is shown in Figure 5b, together with the surface unit cell corresponding to the experimentally observed unit cell in the LEED pattern. As reported by Leicht *et al.* [31], for indistinguishable atoms with similar atomic radii, the LEED pattern is expected to show the orientation and periodicity of the nearest-neighbor cell (black dashed square in Figure 5a, Figure 5b). Instead, the observed surface unit cell is √2 larger (orange dashed square in Figure 5a, Figure 5b), providing support for a surface termination mainly at (Mn, Ga) planes rather than at Ni planes, with a possible buckling of the top surface plane induced by a vertical relaxation of one sort of atoms [31,32].

Upon cooling in front of the LEED across the transition from the austenitic to the martensitic phase, the sharp spots of the quadratic pattern split into a set of fainter reflections in Figure 5c. This complex LEED pattern has been interpreted to be a consequence of the surface topography and the different twin variants appearing upon the martensitic transformation [31]. The austenitic phase has a cubic L2$_1$ structure with $a_{L21}$ = 5.82 Å [17]. The martensitic state has a more complicated 7M modulated monoclinic structure [33], which can be approximately described based on a distorted orthorhombic unit cell having $a_{7M}$ = 6.14 Å, $b_{7M}$ = 5.82 Å, $c_{7M}$ = 5.52 Å, $\gamma_{7M}$ = 92.98° (in the austenitic setting) obtained from the X-ray diffraction analysis shown in Figure 3. At the surface, twin variants expose either (*a*,*b*) or (*b*,*c*) surface unit cells in the X-type twins, each with four simultaneously occurring orientations with respect to the substrate, and tilted at some specific angles with respect to the surface normal. In the case of the Y-type twins, twin variants would expose (*a*,*c*) surface unit cells, and no surface tilting is expected in that case, as the twinning planes are perpendicular to the surface normal (Figure 1). Therefore, the observed LEED patterns with split reflections provide evidence that the microstructure of the sample has been switched to pure X-type after the annealing process.

Further evidence was obtained by the "large-scale" topographic STM image (300 × 300 nm$^2$) in Figure 5d, which shows a roof-like surface morphology extending parallel to MgO[110] and MgO[1$\bar{1}$0] directions. A smaller-scale STM image (150 × 150 nm$^2$) of the roof-like surface morphology is shown in Figure 5e in a derivative mode to enhance the contrast of the atomic steps and defects. The height profile (Figure 5f) measured perpendicularly to the corrugation lines shows a triangular shape, with a peak-to-peak height (in the Z-axis) equal to approximately 1.3 nm and an average periodicity (in the X-axis) of 45 ± 6 nm, from which a misorientation angle $\varepsilon$ ~3.3 ± 0.1° can be deduced. According to previous reports [31,34], each corrugation line is a twin lamella (Figure 5g), each side of which corresponds to a different twin variant. In the case of so-called *a-c* twinning in the 7M martensite, one side of the twin lamella exposes (*a*,*b*) type surface unit cells while the other side exposes (*b*,*c*) type surface unit cells and the misorientation angle is $\varepsilon_{a-c}^{7M} = 45° - tan^{-1}\left(\frac{c}{a}\right) = 3.04°$ (using the lattice parameters reported above), which is in agreement with the value deduced by STM. The misorientation angles are smaller for *a-b* and *b-c* types of twinnings ($\varepsilon_{a-b}^{7M}$ ~ 1.5° and $\varepsilon_{b-c}^{7M}$ ~ 1.5°) and can therefore be disregarded. The *a-c* twinning is preferred due to the close match between $b_{7M}$ and the austenitic lattice parameter, thus limiting the stress. One type of variant in Figure 5e shows additional corrugations (a set of parallel lines perpendicular to the twin lamella). The average spacing between lines (width) is ~ 19.6 ± 0.5 Å, but actual distances are not exactly periodic. Peak-to-peak corrugation between these lines (height) is between 0.4 to 0.8 Å. This extra corrugation is related to the modulated martensite superstructure [31], which implies a displacement of the (110) basal planes in the [1$\bar{1}$0] directions. The period of this modulation (width) is seven basal planes for the 7M phase (according to the aforementioned cell parameters, this value is ~ 20.4 Å), *i.e.* close to the value obtained by STM.

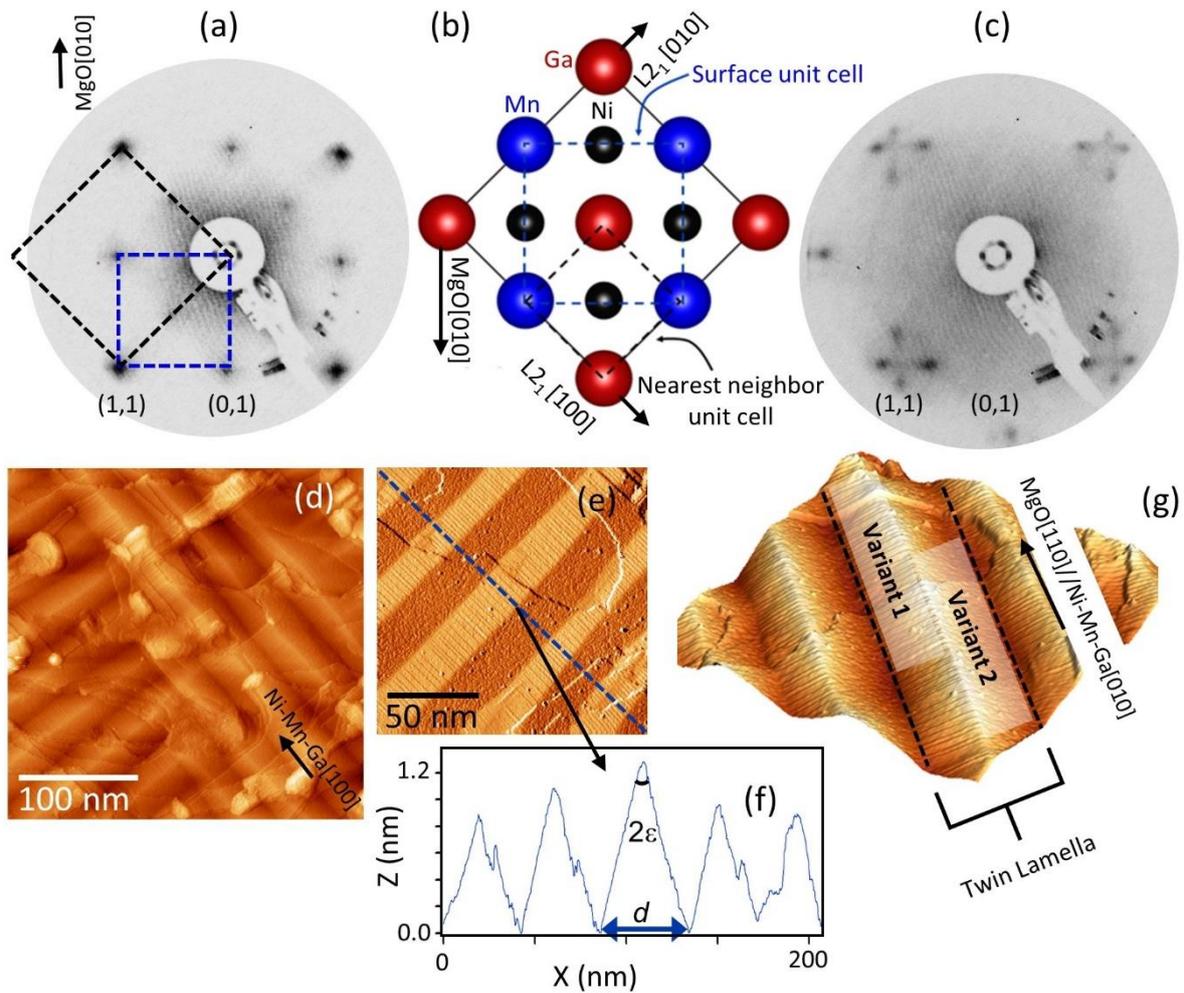

*Figure 5: LEED pattern recorded at 50 eV primary electron beam energy at (a) 348 K and (c) 318 K. (b) Scheme of the austenitic unit cell and its orientation with respect to the MgO substrate. (d) STM topographic image (300 × 300 nm$^2$; $V_b$ = -1.8 V, $I_t$=0.1 nA) of the surface recorded at room temperature after the annealing process. Derivative (e) and 3D view (g) of an STM image of the same surface (150 × 150 nm$^2$, $V_b$ = -1.8 V, $I_t$=0.1 nA). (f) The height profile measured perpendicular to the twin lamellae along the blue line in (e).*

3.3. Microfabrication

On the last step, a piece of the as-grown thin film sample (5 × 5 mm$^2$) was patterned by UV lithography, fabricating arrays of micropatterns including rings, stripes, square-bands, L-stripes, squares, and disks. The martensitic configuration of the fabricated arrays was evaluated before and after the annealing process by a Kerr microscope at room temperature, where different orientations of X-type and Y-type twins interact differently with the polarized light showing different tonality of the gray color [29] (Figure 6). The fabricated arrays before the annealing process show the dominance of Y-type twins, covering a range of 52-88% of the area of the micropatterns (Figure 6a). After the annealing process, the fabricated arrays were recaptured by the Kerr microscope; they show no evidence of the typical gray contrast arising from the Y-type twins (Figure 6b). This observation is in agreement with all the results obtained for the continuous thin film samples suggesting the switch of martensitic configuration to X-type as a result of the annealing process.

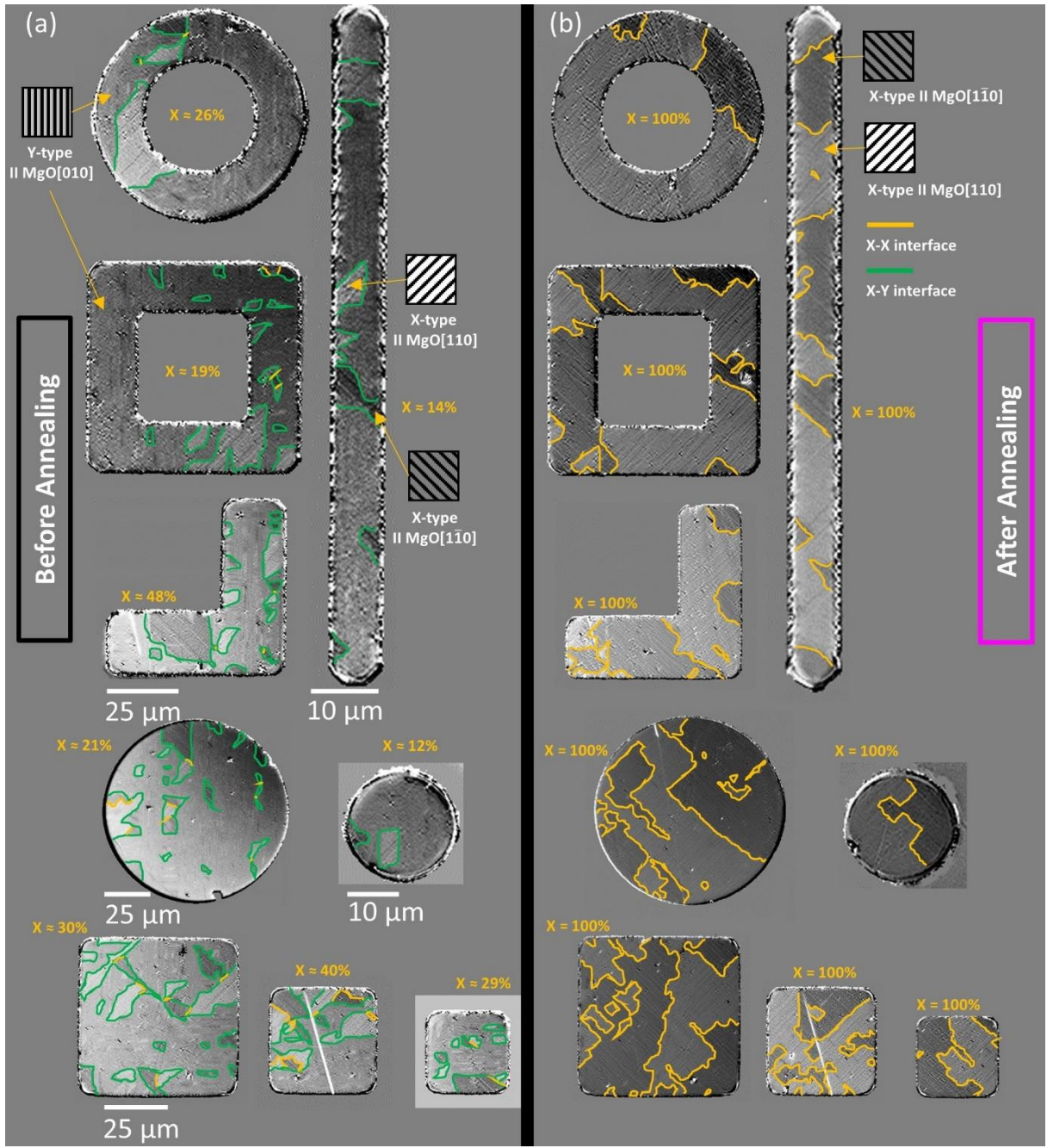

*Figure 6: Microfabricated Ni-Mn-Ga epitaxial structures (a) before and (b) after the annealing process.*

## 4. Discussion

We have reported a simple strategy for promoting the magnetic stray-field contribution at the surface of epitaxially grown Ni-Mn-Ga thin films and spatially confined structures having negligible magnetic stray field contribution: an annealing process was sufficient for switching the hierarchical self-assembly of the martensitic magnetic cells from Y-type into X-type twinning configurations with considerable magnetic stray-field contribution at the surface. Here, we introduce a phenomenological hypothesis, which describes the observed annealing-induced selection of the twins based on the orientations of internal stress in our films.

When epitaxial Ni-Mn-Ga film (lattice parameter ≈ 5.8 Å) is grown on a Cr layer (lattice parameter ≈ 2.9 Å) that is itself epitaxially grown on a MgO(001) substrate (lattice parameter ≈ 4.2 Å), we assume that the primary stress condition mainly originates from the lattice-mismatch at the growth temperature and subsequently the film encounters dynamic thermal stress as a function of temperature because Ni-Mn-Ga, Cr, and MgO have different thermal expansion coefficients [35]. The MgO substrate provides the initial crystallographic orientation but does not impose direct stress on the Ni-Mn-Ga layer due to the buffer effect of the Cr layer. The Cr layer essentially isolates Ni-Mn-Ga from any direct stresses imposed by the MgO substrate. MgO orients the Cr layer such that Cr(001)[110] aligns with MgO(001)[100]. Consequently, the Ni-Mn-Ga film is oriented as Ni-Mn-Ga(001)[110] ∥ Cr(001)[110] (**Figure 7**a). The Ni-Mn-Ga film's in-plane stress is primarily influenced by the Cr layer, which generates biaxial stress along Ni-Mn-Ga[100] and [010] directions (Figure 7b). Martensitic transformation proceeds with a shear-like mechanism, and thus interacts with stress [36]. Moreover, all the possible habit planes and twinning planes in Ni-Mn-Ga cubic cells are close to {110} planes [24] (Figure 7c), which can interact with the stress giving rise to stress-induced selection of twins [24-28]. The resolved shear stress (RSS) on the {110} planes of cubic austenite can be calculated from the following general formula:

*Equation 1*

$$\tau = \sigma \times \cos(\varphi) \times \cos(\lambda)$$

where:

$\sigma$ is the magnitude of the stress, $\varphi$ is the angle between the direction of the stress and the normal to the twinning plane, $\lambda$ is the angle between the direction of the stress and the twinning direction.

The biaxial stress is imposed to cubic Ni-Mn-Ga by epitaxial crystal relation along Ni-Mn-Ga[100] and [010], therefore the stress along Ni-Mn-Ga[100] ($\sigma_1$) equals to the stress along Ni-Mn-Ga[010] ($\sigma_2$).

The twinning direction for each of the {110} planes is equal to the normal direction to that twinning plane (Figure 7c), therefore we have:

(110) with twinning direction along [1$\bar{1}$0], (1$\bar{1}$0) with twinning direction along [110], (101) with twinning direction along [10$\bar{1}$], (10$\bar{1}$) with twinning direction along [101], (011) with twinning direction along [01$\bar{1}$] and (01$\bar{1}$) with twinning direction [011].

The stress is in the plane of the film so, we can consider only the in-plane angles (Figure 7d):

The (110) and (1$\bar{1}$0) planes are at 45° to the stress orientations, so they have $\varphi = 45°$ and $\lambda = 45°$, which gives rise to $\tau = \sigma_1/2 + \sigma_2/2$ for each of these planes. For the other four planes (101), (10$\bar{1}$), (011), and (01$\bar{1}$) however, the angles are $\varphi = 0°$ or 90°, $\lambda = 0°$ or 90°, which gives rise to $\tau = \sigma_2$ for (101) and (10$\bar{1}$), $\tau = \sigma_1$ for (011) and (01$\bar{1}$) planes.

Therefore, the amount of the RSS on each of the six {110} planes is equal but its symmetry is different. In particular, when Ni-Mn-Ga is epitaxially grown on Cr/MgO, each of the four inclined planes (generating X-type twins) accommodate the biaxial stress, asymmetrically ($\tau = \sigma_2$ or $\tau = \sigma_1$); whereas each of the two normal planes (generating Y-type twins) accommodate the biaxial stress, symmetrically ($\tau = \sigma_1/2 + \sigma_2/2$). This could explain our experimental observation where we reported the dominance of Y-type twins at room temperature in our samples before annealing (Figure 2, Figure 3, Figure 6).

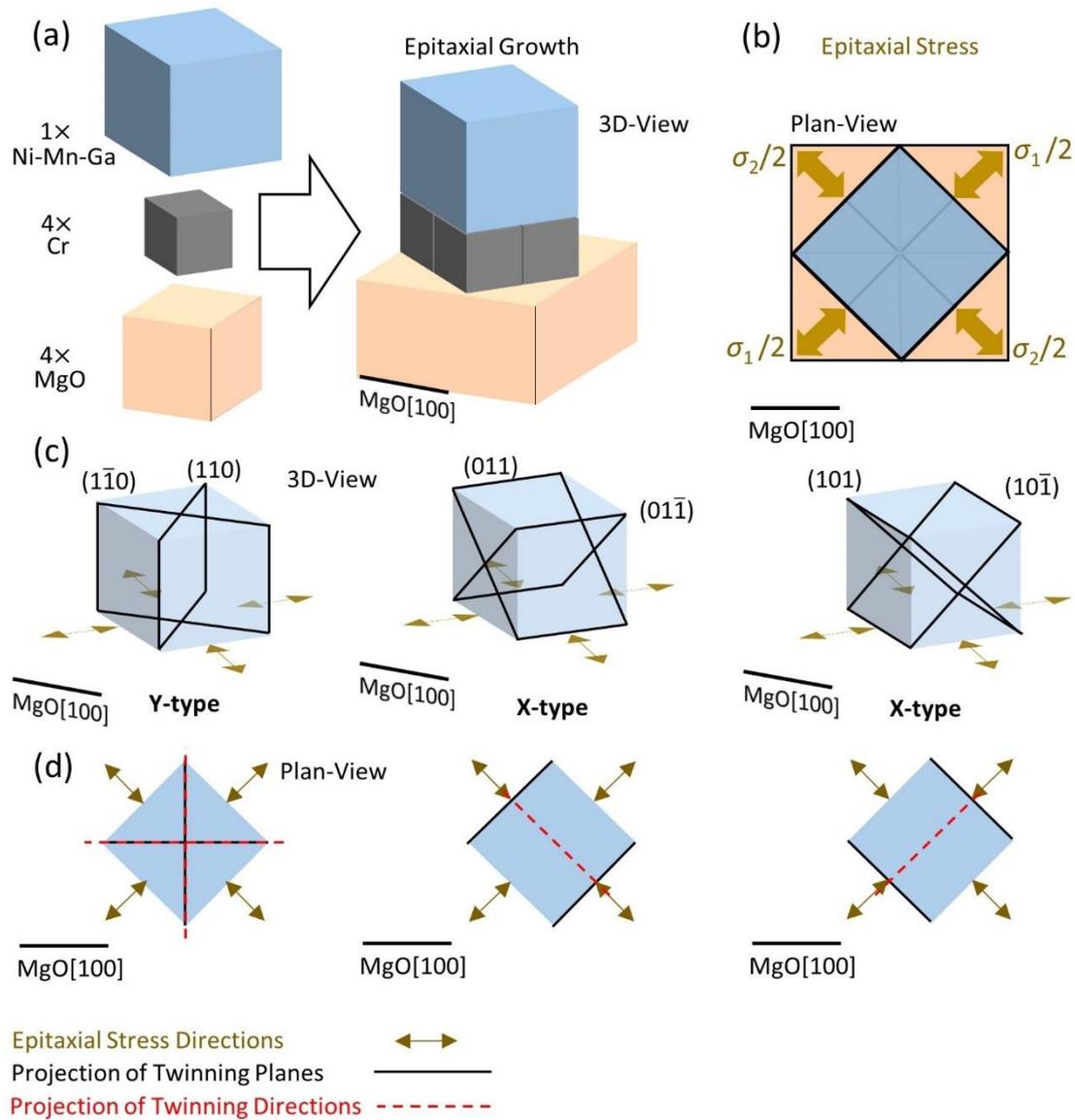

*Figure 7: (a) schematic 3D-view of the epitaxial relation between MgO(001), Cr(001) and Ni-Mn-Ga(001). (b) Plan-view of the scheme shown in (a), highlighting the orientations of the biaxial stress imposed by Cr to Ni-Mn-Ga. (c) 3D-view of the Ni-Mn-Ga cubic austenite, showing the relative orientations between the six possible {110} habit planes (or twinning planes) and the biaxial stress. (d) The relative plan-view of the schemes shown in (c). The red dashed lines are the projections of the twinning directions for each of the {110} planes. Twinning direction for each of the planes is equal to the normal direction of that plane.*

The annealing process involves heating the film to a high temperature and holding it at that temperature. This process allows atoms within the film to diffuse and migrate. Defects in the crystal lattice can move more freely, facilitating stress redistribution or/and relaxation. If the redistribution of the stress in the Ni-Mn-Ga film occurs upon annealing, it may cancel out or modify the primary stress condition, consequently modifying the selection of twins in the martensitic phase. For instance, the redistribution of the primary stress may give rise to an out-of-plane component after the annealing process due to the film's attempt to maintain mechanical equilibrium as the internal stress state changes. In this case, the RSS on the {110} planes of cubic austenite would also change; the (110) and ($1\bar{1}0$) planes would have $\varphi = 90°$ and $\lambda = 90°$ with respect to the out-of-plane component, which gives rise to $\tau = 0$. For the (101), ($10\bar{1}$), (011), and ($01\bar{1}$) planes however, the angles would be $\varphi = 45°$, and $\lambda = 45°$, which gives rise to $\tau = \sigma_{out}/2$ for each of the planes. Therefore, the four inclined {110} planes (generating X-type twins) would be the only planes that can interact with an out-of-plane stress component and this could describe the switching of the twins in our films, where Y-type twins switch into X-type after the annealing process.

To further evaluate our hypothesis, we have performed in-situ annealing for the as-grown sample in the diffractometer, tracing the out-of-plane lattice parameters of the Cr layer and Ni-Mn-Ga film in the process of heating, annealing, and cooling. The results are shown in **Figure 8**, which confirms our hypothesis:

At room temperature, the sample is in the martensitic phase showing the $b(040)$ peak, which corresponds to the Y-type twins (Figure 1). Upon heating, the sample transforms to the austenite phase, the lattice parameters of Cr(002) and Ni-Mn-Ga(004)$_{Aus}$ expand by around 0.003 Å and 0.018 Å, respectively, due to the thermal expansion (Figure 1c, Figure 1e). However, during the annealing process, where the temperature is constant for 60 min, the lattice parameter of Ni-Mn-Ga drops by 0.013 Å while the lattice parameters of Cr is constant, evidencing a strain variation in the out-of-plane direction of the film (Figure 1f). By cooling, the drop of the lattice parameter of Ni-Mn-Ga accelerates in such a way that the Aus(004) peak almost overlaps the Cr(002) peak at $T$ = 380 K (Figure 1d). Finally, by cooling the sample across the phase transition, the martensitic lattice parameter *b*(040) is not visible anymore due to the "switch of the twins" into X-type while the Cr(002) peak is detected in its original position at 298K (Figure 1b). The calculated out-of-plane lattice parameters of Cr and Ni-Mn-Ga are shown in Figure 8g and Figure 8h graphs.

Therefore, we can conclude that the annealing process has modified the stress condition in our films by redistributing the primary biaxial stress imposed by the Cr layer to Ni-Mn-Ga along [100] and [010] directions. This has given rise to an out-of-plane stress variation with which, only the four inclined planes (generating X-type twins) can interact. In other words, the annealing process has imposed new stress conditions in our films and micropatterns, which favors the formation of X-type twinning configurations in the martensitic phase.

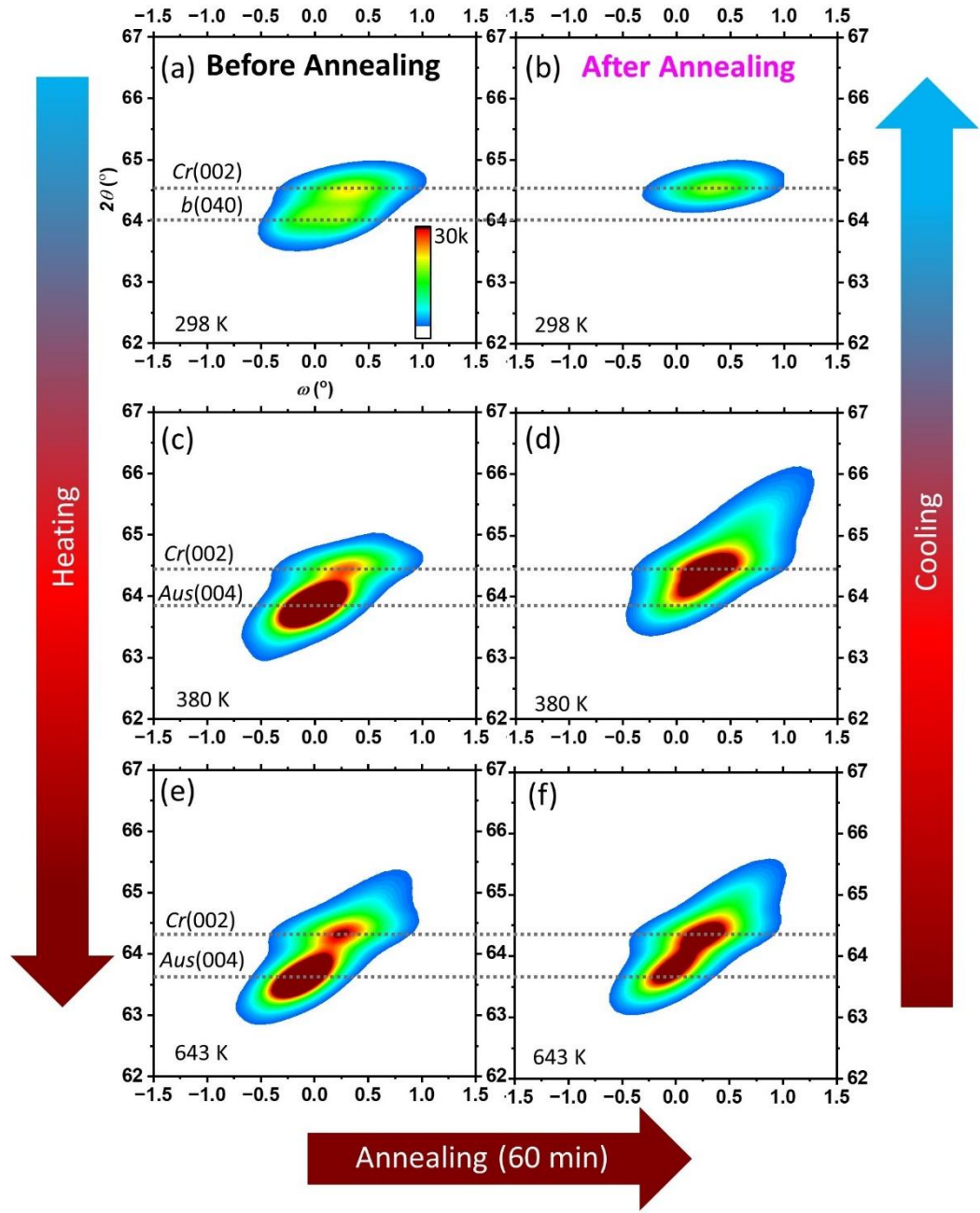
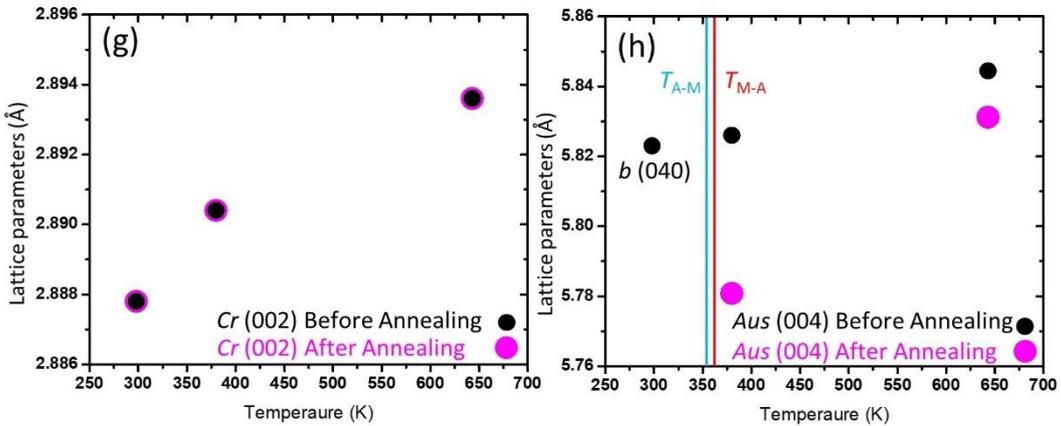

*Figure 8: Out-of-plane 2θ/ω angular maps of the sample (a, c, e) before and (b, d, f) after the annealing process as a function of temperature. (g) Calculated lattice parameter of Cr(002) as a function of temperature. (h) Calculated lattice parameter of Aus(004) as a function of temperature. The black circles refer to the lattice parameters before annealing and pink circles refer to the lattice parameters after annealing (Uncertainty: ~$10^{-3}$Å). The black circle at 298 K refers to b(040) in the martensitic phase.*

## 5. Conclusions

We have introduced a simple strategy for promoting the magnetic stray-field contribution at the surface of epitaxially grown multiferroic Ni-Mn-Ga thin films and spatially confined structures having negligible magnetic stray-field contribution. We reported that an annealing process is sufficient to promote the magnetic stray-field contribution at the surface by switching the martensitic twinning configurations from Y-type to X-type. All the structural, morphological, and surface characterizations supported our conclusion. Switching of twins was evidenced after the annealing process by the absence of the bright contrast in BSE micrographs, the change in the intensity pattern of the peaks in the X-ray maps, splitting of the LEED diffraction spots at the martensitic transition temperature (~338 K) upon cooling, and the absence of the Y-type contrast in Kerr micrographs. The surface microstructure of the sample after the annealing process, its roof-like morphology, and all associated corrugation features were specific to *a-c* twinning of the 7M martensitic structure and indicative of the X-type twinning configuration. The magnetic evolution of the sample after the annealing process was evidenced in the MFM micrographs by the dominance of the magnetic stray field arising from the perpendicular anisotropy contribution. The magnetic evolution was also evidenced by the significant reduction of the remanence, coercivity, and the absence of the typical Y-type jumps in the magnetization loops of the sample measured after the annealing process.

We have discussed the annealing-induced "switching of twins" by introducing a phenomenological hypothesis, which describes the twinning selection based on the directions of internal stress in our films: when Ni-Mn-Ga is epitaxially grown on Cr/MgO, the two normal planes (generating Y-type twins) are the only {110} available planes that can symmetrically accommodate the biaxial stress imposed by Cr layer along Ni-Mn-Ga[100] and [010]. This condition is modified after the annealing process; it imposes a new stress condition in our films and micropatterns, which favors the formation of X-type twinning configurations in the martensitic phase.

**Acknowledgements**

This work was financed by the European Union - NextGenerationEU (National Sustainable Mobility Center CN00000023, Italian Ministry of University and Research Decree no. 1033, 17/06/2022, Spoke 11, Innovative Materials & Lightweighting). The opinions expressed are those of the authors only and should not be considered as representative of the European Union or the European Commission's official position. Neither the European Union nor the European Commission can be held responsible for them. This Project was funded by European Union – NextGenerationEU under the National Recovery and Resilience Plan (NRRP), Mission 4 Component 2 Investment 1.1 - Call for tender No. 1409 of 14-09-2022 of Italian Ministry of University and Research (Project Code P2022KMXBL, Concession Decree No. 0001381 of 01/09/2023 adopted by the Italian Ministry of Universities and Research, CUP D53D23019360001, "Small-scale Thermomagnetic Energy harvesters: from materials to devices"). This work was supported by the project CZ.02.01.01/00/22_008/0004594. Access to the CEITEC Nano Research Infrastructure was supported by the Ministry of Education, Youth and Sports (MEYS) of the Czech Republic under the project CzechNanoLab (LM2023051). VKS, JL and VF acknowledge the support from « Lorraine Université d'Excellence », part of the France 2030 Program, ANR-15-IDEX-04-LUE and the European Integrated Center for the Development of Metallic Alloys and Compounds (ECMetAC). VF benefited from the support of CNR (prot. n°0262747).